\documentclass[journal]{IEEEtran}
\usepackage[T1]{fontenc} 

\usepackage{cite}

\usepackage[hang,flushmargin]{footmisc}

\usepackage{graphicx}
\usepackage{amsmath}
\usepackage{mathtools}
\usepackage{comment}
\usepackage{commath}
\usepackage{enumitem}
\usepackage{booktabs}
\usepackage{multirow} 
\usepackage{epstopdf}
\usepackage{array}
\usepackage{wrapfig}

\newcommand{\innermid}{\nonscript\;\delimsize\vert\nonscript\;}
\newcommand{\activatebar}{%
  \begingroup\lccode`\~=`\|
  \lowercase{\endgroup\let~}\innermid 
  \mathcode`|=\string"8000
}

\usepackage{amsthm,amssymb}
\usepackage{setspace}

\newcommand{\subparagraph}{}
\usepackage[compact]{titlesec}

\usepackage{url}
\usepackage[breaklinks]{hyperref}

\hypersetup{colorlinks=true,
            linkcolor=blue,urlcolor=blue,
            anchorcolor=blue,citecolor=blue}
            
\usepackage{breakurl}

\usepackage{bm}
\usepackage{tikz}
\usetikzlibrary{shapes,arrows}

\usepackage{caption}
\usepackage{array, tabularx}
\usepackage[newcommands]{ragged2e}

\begin{document}

\tikzstyle{decision} = [diamond, draw, fill=blue!20, 
    text width=4.5em, text badly centered, node distance=3cm, inner sep=0pt]
\tikzstyle{block} = [rectangle, draw, fill=blue!20, 
    text width=5em, text centered, rounded corners, minimum height=4em]
\tikzstyle{line} = [draw, -latex']
\tikzstyle{cloud} = [draw, ellipse,fill=red!20, node distance=3cm,
    minimum height=2em]


%
\title{SpatioTemporal Feature Integration and Model Fusion for Full Reference Video Quality Assessment}
%
%
%

\author{Christos G. Bampis, Zhi Li and Alan C. Bovik
\thanks{C. G. Bampis, and A. C. Bovik are with the Department
of Electrical and Computer Engineering, University of Texas at Austin, Austin,
USA (e-mail: bampis@utexas.edu; bovik@ece.utexas.edu). Z. Li is with Netflix Inc. (e-mail: zli@netflix.com).}
}

\maketitle

\begin{abstract}

Perceptual video quality assessment models are either frame-based or video-based, i.e., they apply spatiotemporal filtering or motion estimation to capture temporal video distortions. Despite their good performance on video quality databases, video-based approaches are time-consuming and harder to efficiently deploy. To balance between high performance and computational efficiency, Netflix developed the Video Multi-method Assessment Fusion (VMAF) framework, which integrates multiple quality-aware features to predict video quality. Nevertheless, this fusion framework does not fully exploit temporal video quality measurements which are relevant to temporal video distortions. To this end, we propose two improvements to the VMAF framework: SpatioTemporal VMAF and Ensemble VMAF. Both algorithms exploit efficient temporal video features which are fed into a single or multiple regression models. To train our models, we designed a large subjective database and evaluated the proposed models against state-of-the-art approaches. The compared algorithms will be made available as part of the open source package in \url{https://github.com/Netflix/vmaf}.


\end{abstract}


%
\IEEEpeerreviewmaketitle

\section{Introduction}
%
%
%
%

\IEEEPARstart{V}\ \hspace{-0.89mm}\MakeLowercase{i}deo traffic from content delivery networks is expected to rise to 71\% by 2021 \cite{cisco-vni}. For numerous video applications, such as adaptive video streaming, consumer video applications and digital cinema, perceptual video quality assessment (VQA) is an integral component. The enormous amount of streaming (YouTube or Netflix) and social media (Snapchat, Instagram, Facebook, among many) video data emphasizes the need for measuring and controlling video quality. In practice, objective video quality prediction models are frequently deployed for quality assessment. VQA models can also be useful for conducting codec comparisons \cite{de2016large,tan2016video}, optimizing perceptual video coding protocols \cite{wang2016ssim} or as inputs to quality of experience (QoE) predictors \cite{duanmu2017quality, VATL, 7931662}.

VQA models can be classified into three main categories \cite{chikkerur2011objective}: full-reference (FR), reduced-reference (RR) and no-reference (NR) models. FR VQA models require an entire reference video signal to measure visual quality \cite{wang2004video,sheikh2005visual,vu2011spatiotemporal,pinson2014temporal,manasa2016optical,seshadrinathan2010motion, techblog}, while RR approaches exploit a limited amount of reference information \cite{wang2005reduced,li2009reduced,ma2011reduced,rehman2012reduced,soundararajan2013video,bampis2017speed}. NR models exploit distortion-specific or natural video statistical models to predict quality without using any reference information \cite{mittal2016completely,zhang2015feature,saad2014blind,li2016spatiotemporal,chen2016perceptual,xu2014no,zhu2013no,lin2012no,sogaard2015no}. The focus of our work is FR VQA models.

There have been numerous approaches to the design of FR VQA algorithms. Image-based approaches \cite{wang2004image,sheikh2006image} exploit only spatial information by capturing statistical and structural irregularities between distorted video frames and corresponding reference frames. A common principle underlying many of these models is that bandpass-filtered responses of high-quality video frames can be modeled as Gaussian Scale Mixture (GSM) vectors \cite{wainwright1999scale,portilla2003image} and that distorted frames can be quantified in terms of statistical deviations from the GSM model. The GSM approach has been applied in the spatial \cite{wang2004image,bampis2017speed}, wavelet \cite{soundararajan2013video,sheikh2005visual} and DCT \cite{saad2014blind} domains. Importantly, frame differences of high quality videos can also be modeled as GSM vectors in order to measure temporal video distortions \cite{sheikh2005visual,soundararajan2013video,bampis2017speed}. 

Video-based FR VQA models have also been studied in the literature \cite{vu2011spatiotemporal,pinson2014temporal,manasa2016optical,seshadrinathan2010motion}. In \cite{vu2011spatiotemporal}, the notion of spatio-temporal slices was derived and the ``most apparent distortion principle'' \cite{larson2010most} was applied to predict video quality. Optical flow measurements were also used in \cite{manasa2016optical}, where video distortions were modeled by deviations between optical flow vectors. A space-time Gabor filterbank was used in \cite{seshadrinathan2010motion} to extract localized spatio-spectral information at multiple scales. VQM-VFD \cite{pinson2014temporal,pinson2004new} used a neural network trained with a large number of features such as edge features. These algorithms often deliver good performance on small size videos, but are computationally inefficient on long HD video sequences, since they apply time-consuming spatio-temporal filtering operations.

Data-driven models hold great promise for the VQA problem \cite{techblog,pinson2014temporal,mocanu2015no,xu2014no,le2006convolutional,lin2014fusion,lin2015evqa}. Netflix recently announced the Video Multimethod Fusion Approach (VMAF), which is an open-source, learning-based FR VQA model. VMAF combines multiple elementary video quality features using an SVR trained on subjective data, and focuses on compression and upscaling artifacts. Nevertheless, it does not fully exploit temporal quality information sensitive to temporal video distortions.

The open-sourced VMAF framework can be used as a starting point to develop better VQA models by integrating stronger quality features and training data. Here we leverage these capabilities by proposing two ways to improve upon the current VMAF framework. The first approach, called SpatioTemporal VMAF, integrates strong temporal features into a single regression model. The second enhancement (Ensemble VMAF) trains two separate models and then performs prediction averaging to predict video quality. Both approaches rely on statistical models of frame differences and hence avoid computationally expensive spatio-temporal filtering. To train our models, we designed a large subjective experiment (VMAF+ database) and evaluated these models in three experimental applications: video quality prediction, QoE prediction, and Just-Noticeable Difference (JND) prediction.

The rest of the paper is organized as follows. Section \ref{VMAF} describes the current VMAF system and highlights its capabilities and limitations. Sections \ref{STVMAF} and \ref{EVMAF} respectively discuss the SpatioTemporal and Ensemble VMAF improvements. Section \ref{VMAF+} gives an overview of the VMAF+ subjective dataset that we built. Section \ref{experiments} details experimental results, while Section \ref{Future} concludes with ideas for future work.

\section{Background on VMAF}
\label{VMAF}

\subsection{How VMAF works}

\begin{figure}[htp]
\centering
\captionsetup{justification=centering}
\includegraphics[width = \columnwidth]{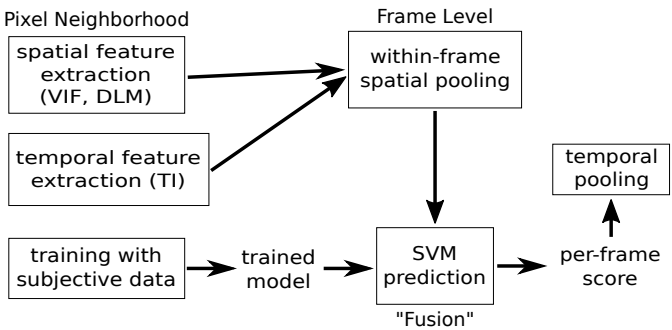}
\caption{Outline of the current VMAF system.}
\label{current_VMAF}
\end{figure}

VMAF extracts a number elementary video quality metrics as features and feeds them into an SVR \cite{techblog}. This allows VMAF to preserve and weight the strengths of each individual feature and align the objective predictions with ground truth subjective data. The VMAF system includes the following steps (see Fig. \ref{current_VMAF}): feature extraction and aggregation, training/testing and temporal pooling.

The first step is to extract a number of quality metrics as perceptually-relevant features: DLM \cite{li2011image}, VIF \cite{sheikh2006image} and the luminance differences between pairs of frames (Temporal Information - TI). The DLM feature captures detail losses and is calculated by a weighted sum of DLM values over four different scales. The VIF feature captures losses of visual information fidelity and is computed at four scales, yielding four VIF features. The TI feature aims to capture temporal effects due to motion changes which are quantified by luminance differences, resulting in six features overall. The TI feature is currently the only source of temporal quality measurement in VMAF.

Each of these six features is extracted as a feature map of size equal to the corresponding scale. Next, the average value of each feature map is calculated, to produce one feature value per video frame and feature type. For training purposes, VMAF aggregates the per frame features over the entire video sequence, yielding one feature value per training video. These six feature values are fed, together with the corresponding subjective ground truth, to an SVR model. For testing purposes, VMAF predicts one value per video frame and calculates the arithmetic mean over all per frame predictions to predict the overall video quality.

\subsection{VMAF Limitations and Advantages}

VMAF has been developed with a particular application context in mind. For the Netflix use case, there are two main video impairments that are of interest: compression and scaling artifacts. Compression artifacts are typically observed as blocky regions within a frame, while scaling artifacts arise when the encoding resolution is lower than the display resolution and are usually observed as jerky regions around edges. Both of these artifact types are introduced while encoding the video content. Packet loss transmission distortions are not a problem for HTTP adaptive streaming applications which rely on the TCP transfer protocol. 

\begin{figure*}[htp]
\centering
\includegraphics[width = 0.5\columnwidth]{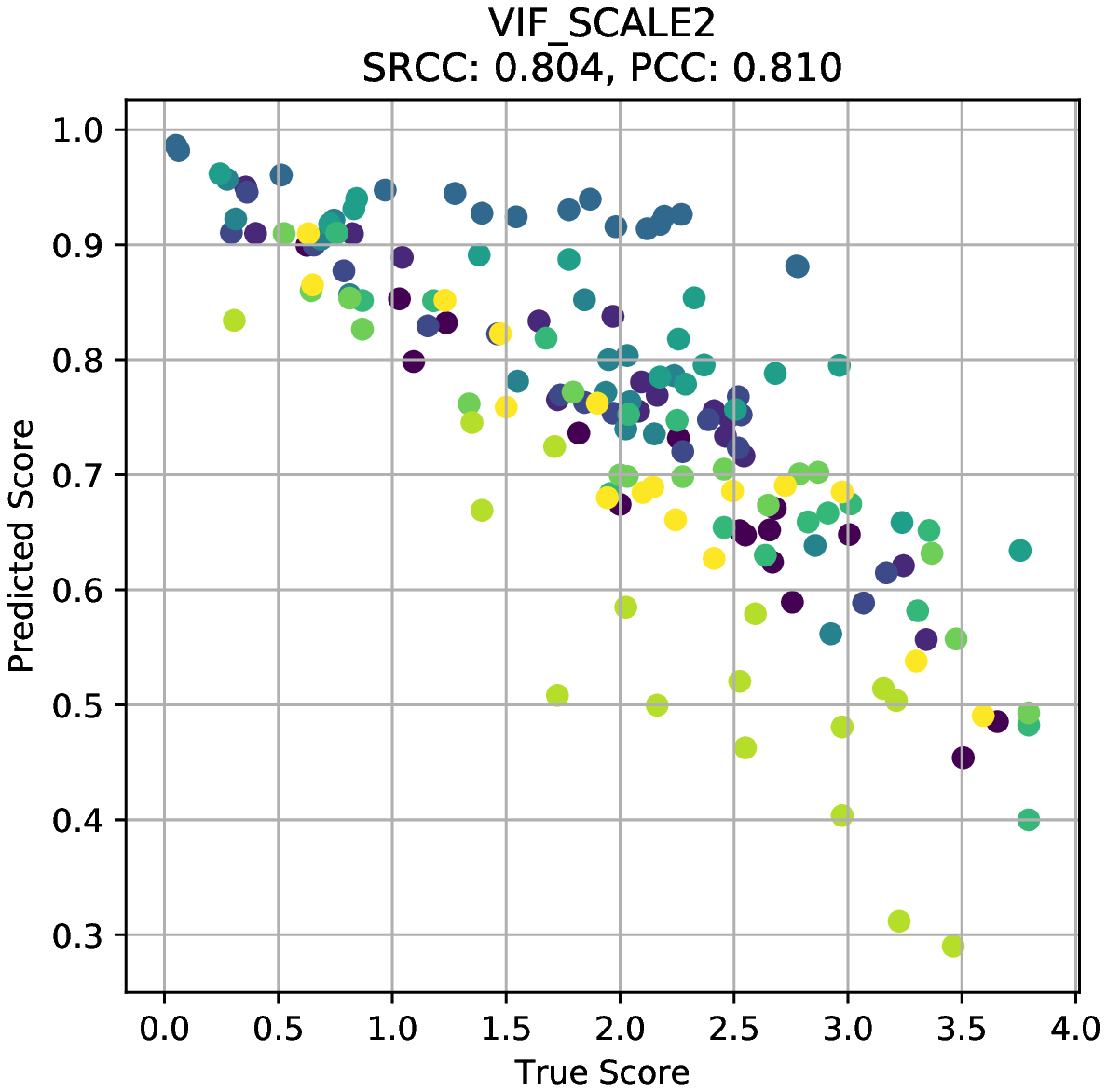}
\includegraphics[width = 0.5\columnwidth]{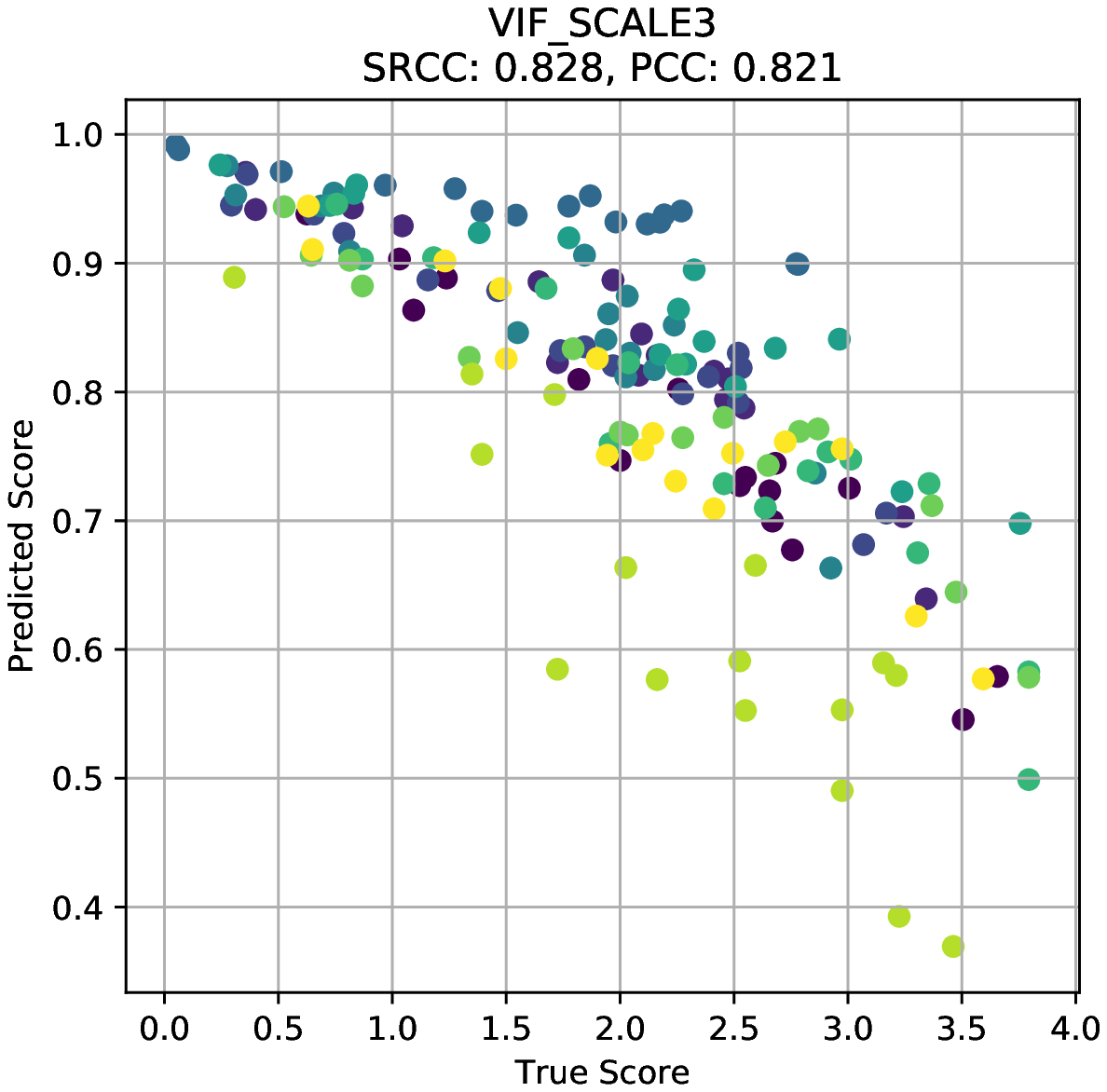}
\includegraphics[width = 0.5\columnwidth]{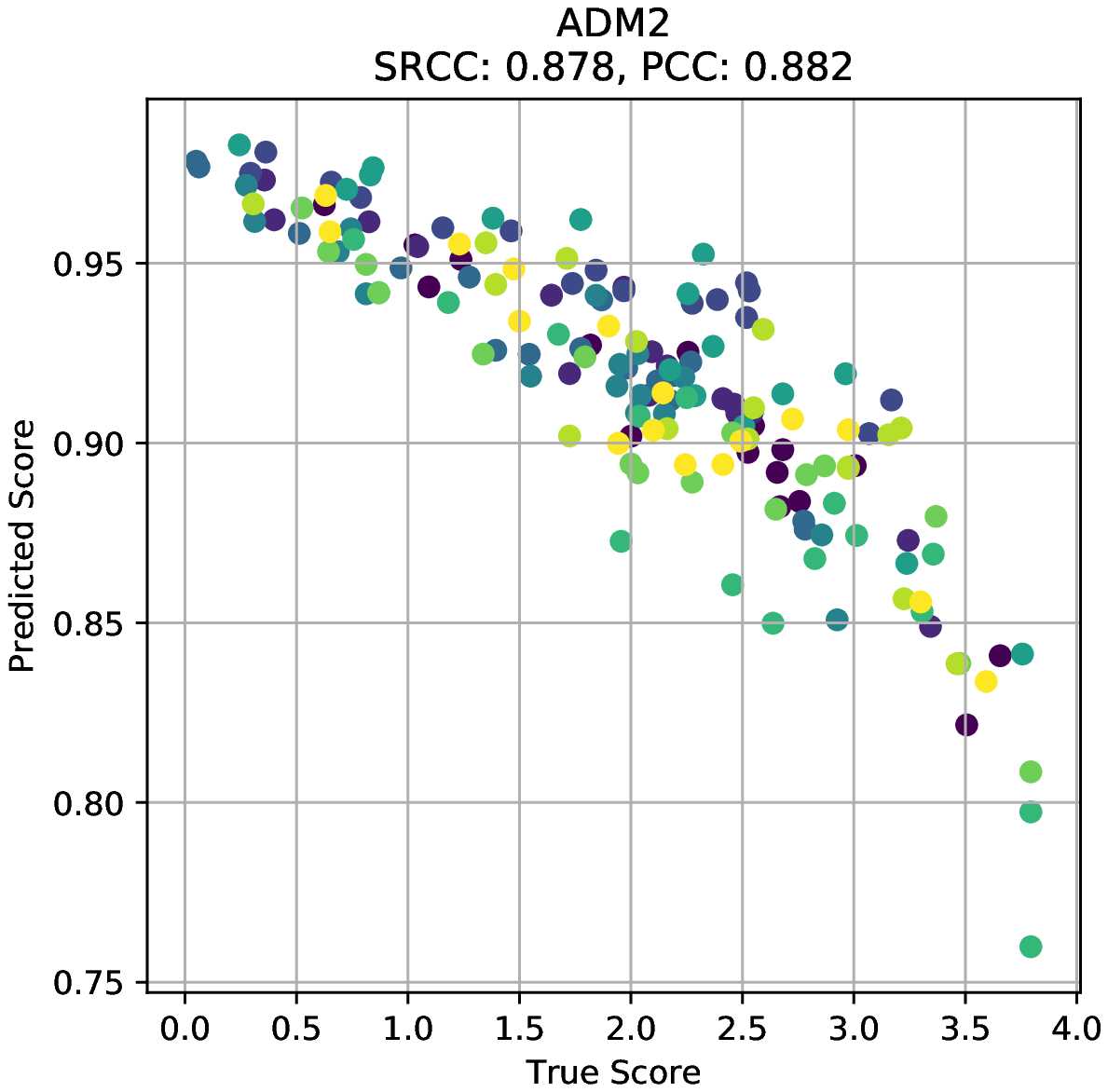}
\includegraphics[width = 0.5\columnwidth]{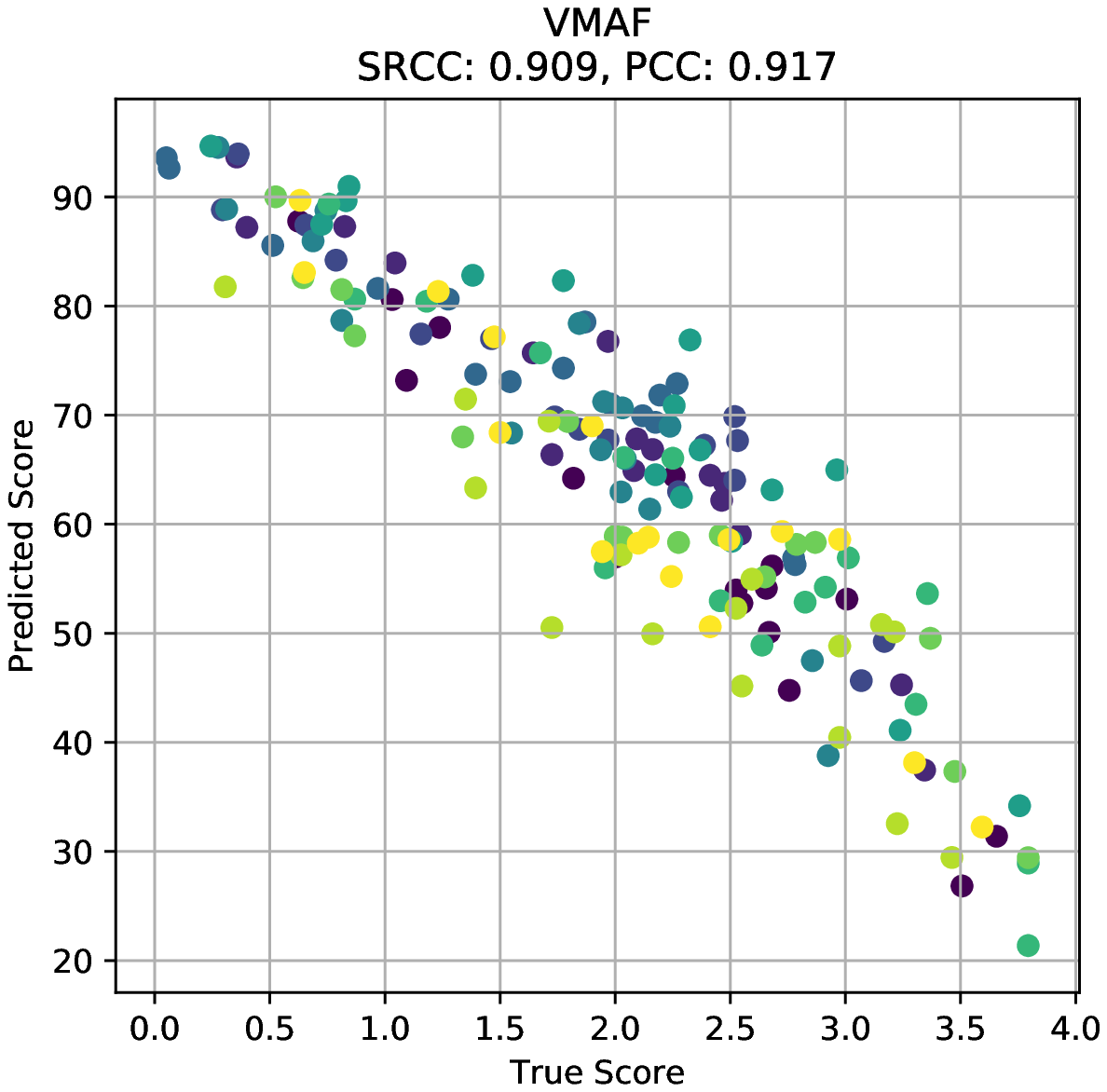}
\caption{Performances of the individual VMAF features and the fusion result on the LIVE Mobile VQA Database \cite{moorthy2012video}. Left to right: VIF calculated at scales 2 and 3; DLM; VMAF fusion. When training VMAF, we relied on the NFLX dataset \cite{techblog}. The performance metrics and our model evaluation are described in greater detail in Section \ref{experiments}.}
\label{elementary}
\end{figure*}

Under this specific application context, VMAF achieves good predictive performance by weighting the elementary video quality features. Figure \ref{elementary} illustrates an example of the performance gains afforded by VMAF fusion. Importantly, VMAF has been trained on video sources which contain film grain noise. In practice, film grain may be found in older (legacy) content, but also in newer content, where film grain is synthetically added with an artistic intent. The effects of film grain on perceived video quality are not always clear, since film grain may be reduced due to compression and sometimes possesses an aesthetic subjective appeal. By training on the presence of film grain, VMAF ``learns'' to account for these phenomena when performing video quality predictions.

\begin{figure*}[htp]
\centering
\includegraphics[width = 1.8\columnwidth]{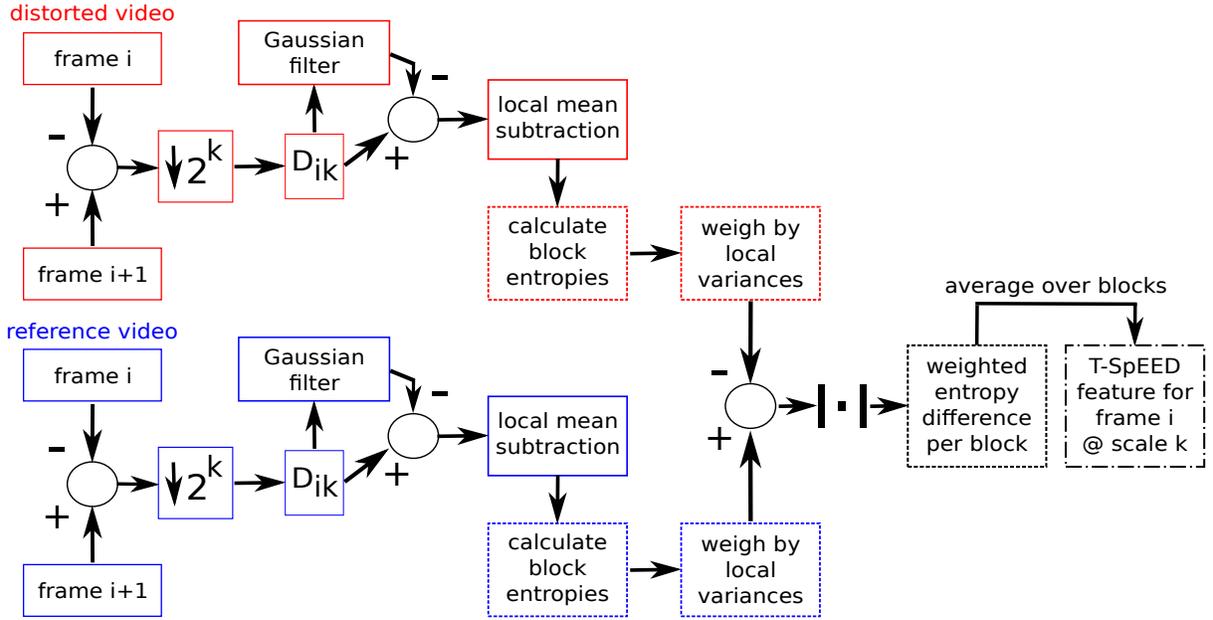}
\caption{T-SpEED feature extraction. Blue and red colors denotes the reference and distorted videos respectively. A dashed box outline denotes that these operations are performed on each block of the MS map, while dashed and bulleted outline denotes a single value per frame. When extracting the S-SpEED features, the diagram remains the same, except that whole video frames are used instead of frame differences.}
\label{feature_extraction}
\end{figure*}

Aside from average frame difference measurements (TI feature), VMAF does not exploit temporal quality measurements. The TI feature attempts to capture motion masking effects, i.e., the reduction of distortion visibility due to large motion changes. Nevertheless, TI measurements are more related to the video content itself and do not effectively account for temporal masking. Temporal video distortions, such as ghosting, flickering and motion estimation errors, are quite complex in nature and deeply impact perceived video quality \cite{seshadrinathan2010motion}. Since compression standards are evolving and even lower encoding rates are being used \cite{netflix_low_bitrates}, it is important for FR VQA models, such as VMAF, to generalize well on unseen video distortions.

\section{SpatioTemporal VMAF}
\label{STVMAF}

\subsection{S-SpEED and T-SpEED features}

Extracting temporal quality information is important for VQA models, but space-time VQA models are often computationally intensive, since they employ motion estimation or spatiotemporal filtering. To extract temporal quality measurements, we exploit statistical models of frame differences in high-quality videos similar to \cite{sheikh2005visual,soundararajan2013video} and \cite{bampis2017speed}. The main idea is to model the bandpass-filtered map responses of frames and frame differences as GSM vectors \cite{wainwright1999scale,portilla2003image} and use entropic differencing to predict visual quality. To calculate these entropy values, a conditioning step is applied which removes local correlations from band-pass filtered coefficients. Conditioning is equivalent to divisive normalization \cite{ruderman1994statistics}; a process that is known to occur in the early stages of vision \cite{mittal2012no,carandini1997linearity,field1987relations}.

We build our work on the recently developed SpEED-QA model \cite{bampis2017speed}, which extracts information-theoretic information in the spatial domain. A diagram of the feature extraction steps is shown in Fig. \ref{feature_extraction}. First, let $F_{i}$ be the $i$th video frame and $D_{i}=F_{i+1}-F_{i}$ be the $i$th frame difference of the reference or the distorted video. Then, downsample $D_{i}$ to the $k$th scale, which yields the $D_{i,k}$ frame difference map. Then filter $D_{i,k}$ with a spatial Gaussian filter and perform local mean subtraction in the spatial domain. This local operation approximates the multi-scale multi-orientation steerable filter decomposition used in \cite{soundararajan2013video} and is very compute-efficient.

\begin{figure}[htp]
\centering
\includegraphics[width = \columnwidth]{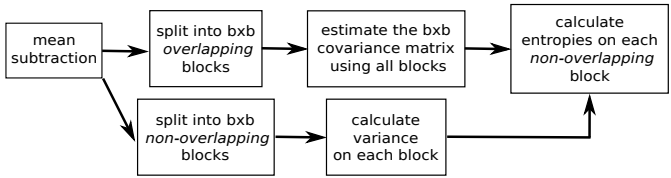}
\caption{Details on entropy calculation for S-SpEED and T-SpEED.}
\label{entropy_calculation}
\end{figure}

Entropy measurements and entropic differencing have been shown to correlate quite highly with human judgements of video quality \cite{bampis2017speed,soundararajan2013video}. Therefore, our next step is to calculate the local entropies in the reference and the distorted video for the local mean-subtracted response map (MS map). These steps are visualized in Fig. \ref{entropy_calculation}. We split the response map into $b\times b$ non-overlapping blocks yielding the coefficients $C_{mk}$ for block $m$ and scale $k$. These coefficients can be modeled as a GSM vector, i.e., $C_{mk} = S_{mk}U_{mk}$, where $S_{mk}$ represents the variance field and is a non-negative random variable independent of $U_{mk}\sim \mathcal{N}(0, \mathbf{K}_{U_{k}})$. We model the neural noise present along the visual pathway using an additive white noise model, i.e, $C'_{mk} = C_{mk} + W_{mk}$, where $W_{mk}\sim \mathcal{N}(0, \sigma_w^2\mathbf{I}_N)$, $\mathbf{I}_N$ is the $b\times b$ identity matrix and $N=b^2$ is the number of coefficients per block. In our implementation, we fix $\sigma_w^2=0.1$, $b=5$ and use a $7\times 7$ isotropic gaussian filter of standard deviation $7/6$.

To predict video quality, SpEED-QA calculates the entropy differences between a reference and a distorted video at the lowest scale ($k=4$). To this end, we also apply conditioning on the block variances $s_{mk}$, which are realizations of $S_{mk}$ and compute the entropies of the noisy bandpass coefficients $C'_{mk}$, i.e.,
\begin{equation}\label{entropy_operation}
h(C'_{mk}|S_{mk}=s_{mk}) = \frac{1}{2}\log [ ( 2 \pi e)^N |s^2_{mk}\mathbf{K}_{U_{k}}+\sigma_w^2\mathbf{I}_N|]
\end{equation}
To determine $s_{mk}$, calculate the sample variance on every non-overlapping block of the MS map. To estimate the $b\times b$ covariance matrix $\mathbf{K}_{U_{k}}$, we use a sliding window to collect all \textit{overlapping} blocks from the MS map and compute the sample covariance. The use of overlapping blocks in this step ensures that a sufficient number of samples is available for covariance estimation, especially for lower scales.

Following entropy calculation, the block entropies are further weighted by a logarithmic factor, i.e., $\log(1+s^2_{mk})$. This step lends a local nature to the model and ensures numerical stability at regions of low spatial or temporal variance. To measure the statistical distance between the GSM models of the distorted and reference video frames, the weighted block entropy values are differenced and the absolute values of those differences are computed. The absolute values are averaged over all blocks, yielding the T-SpEED feature for frame $i$ and scale $k$. This feature captures the information loss due to temporal video distortions. To capture spatial quality degradations, we can also define the corresponding S-SpEED feature by performing local mean subtraction on $F_{i,k}$ instead of $D_{i,k}$, then following the exact same steps.

\subsection{SpatioTemporal Feature Integration}

Despite the good performance of SpEED-QA in a number of databases \cite{bampis2017speed}, it does not account for the effects of film grain on the perceived visual quality. Unlike VMAF, SpEED-QA is not data-driven. Instead, it computes a statistical distance between the best-fitting GSM approximations of the distributions of a reference and a distorted video, rather than modeling the effects of noise present in the source. As we demonstrate in our experimental analysis (see Table \ref{train_vmaf_plus}), SpEED-QA lags in performance on the NFLX dataset, which contains a number of video sources with film grain.

Another shortcoming of SpEED-QA is that it does not exploit multiscale information. Previous studies have established the merits of multiscale information for image and video quality assessment \cite{wang2003multiscale}. The human visual system processes visual information in a multiscale fashion, while images demonstrate significant self-similarities. Notably, multiscale algorithms incorporate the effects of different display sizes and viewing distances. Unfortunately, unlike image quality applications, incorporating multiscale information for VQA is not as easy. 

In preliminary experiments, we discovered that the use of temporal entropy differences across multiple scales yields complementary perceptual information. To exploit this observation and combine information across scales, we adopt a data-driven approach to learn the contribution from each scale and predict visual quality. Due to a motion downshifting phenomenon \cite{soundararajan2013video}, lower scales yield stronger features, hence we extract T-SpEED features from scales 2, 3 and 4. The use of scale $k$ denotes that the frame difference MS map is downscaled by a factor of $2^k$, which allows for more efficient feature extraction.

To complement the T-SpEED features, we found that applying VIF on the frame difference signal \cite{sheikh2005visual} across multiple scales leads to further improved performance. We call these features T-VIF (4 features calculated from scales 0, 1, 2 and 3). Both T-SpEED and T-VIF measure temporal information loss using the GSM statistical model \cite{wainwright1999scale} on frame differences, but T-VIF relies on mutual information between wavelet coefficients. Since the 5 spatial VMAF features (DLM and VIF from 4 scales) sufficiently capture spatial quality degradations, we include them in our model as well. Overall, the proposed SpatioTemporal VMAF (ST-VMAF) approach deploys 12 perceptually relevant features (5 from VMAF, 3 from T-SpEED and 4 from T-VIF) which capture both spatial and temporal information. Compared to other feature candidates, we found that the proposed feature set delivers the best performance.

Similar to the original VMAF approach, we average the per frame features during training but perform per frame ST-VMAF predictions when testing. This design choice did not have an effect on the predictive performance of the ST-VMAF model. This also enables ST-VMAF to be used as an input to a larger, online, QoE prediction system (see Section \ref{experiments}).

To calculate the aggregate quality over an entire video sequence, we applied the hysteresis temporal pooling method in \cite{seshadrinathan2011temporal}. Human opinion scores vary smoothly over time, while objective predictions respond sharply to visual changes. Meanwhile, subjective quality perception is driven by memory/recency, i.e., more recent experiences tend to more deeply affect current visual impressions. Based on these observations, we applied a linear low-pass operator and a non-linear rank order weighting on the objective prediction scores, as suggested in in \cite{seshadrinathan2011temporal}.

\section{Ensemble VMAF}
\label{EVMAF}

\subsection{Why an Ensemble Model?}

In the previous section, we described a simple way to integrate strong temporal quality measurements into VMAF, by concatenating the spatial VMAF features with the T-VIF and T-SpEED features. However, in cases where the available subjective video data is limited, increasing the feature dimensionality (or using deep neural networks) may lead to overfitting. Video databases are usually pretty diverse in their design and contents and hence a particular feature subset may work well on one dataset, but not on another. For example, we have empirically observed that the ADM feature carries a large weight for spatial degradations, but does not generalize well on unseen data. One option is to carefully tune the regression model parameters to effectively regularize the predictions. Another alternative, which we have decided to follow here, is to consider fusion approaches.

Model fusion (or ensemble learning) is a well-studied concept \cite{1688199} which combines multiple individual learners. The main idea is to fuse multiple simple models that do not overfit, are easier to tune, and that complement each other towards reducing the prediction variance. Among other fusion possibilities, we experimented with training multiple SVRs on different video databases or training different regressors (e.g. a Random Forest and a SVR) on the same dataset. Nevertheless, we found that aligning predictions coming from models that were trained on subjective data collected under different experimental conditions and/or assumptions was a difficult proposition. We also found that the SVR predictions always outperformed Random Forest predictions and hence their combination was not beneficial. The performance merits of using an SVR for image and video quality assessment have also been demonstrated in \cite{mittal2012no, techblog, saad2014blind}. These observations led us to the design of Ensemble VMAF (E-VMAF), which we describe next. 

\subsection{An Ensemble Approach to Video Quality Assessment}

\begin{figure}[htp]
\vspace{-4mm}
\centering
\captionsetup{justification=centering}
\includegraphics[width = \columnwidth]{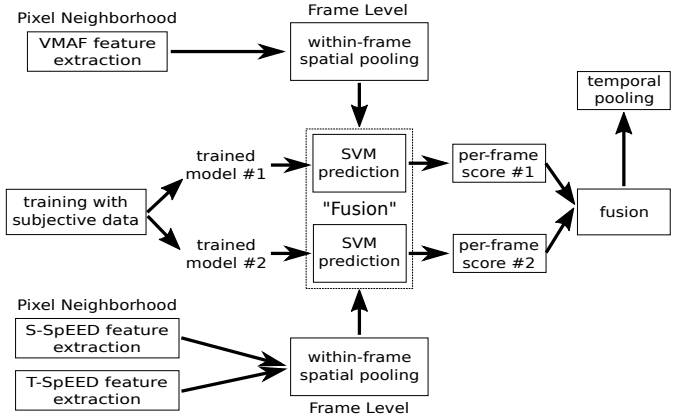}
\caption{Overview of the ensemble approach.}
\label{EVMAF_example}
\end{figure}

We propose E-VMAF, an ensemble enhancement to VMAF, wherein multiple feature subsets are used to train diverse VQA models that are then aggregated to deliver a single prediction value. Nevertheless, training and combining multiple models can significantly increase the complexity, which can be challenging for a VQA model if it is to be deployed at a global scale. Driven by simplicity, we trained two SVR models on the VMAF+ database (see also Section \ref{VMAF+}), and then averaged the individual predictions, as shown in Fig. \ref{EVMAF_example}.

Given that the VMAF feature set already captures the combined effects of compression and scaling (and the predominance of these distortions in practice), we use the same features (6 features) for the first individual model, denoted by M$_1$. To design M$_2$, it is desirable to capture both spatial and temporal quality measurements, such that the individual predictions are accurate enough. Motivated by the perceptual relevance of the T-SpEED features used in ST-VMAF, we combined the 3 T-SpEED features with the 3 S-SpEED features calculated at the same scales (2, 3 and 4). 

The VIF features of M$_1$ and the S-SpEED features of M$_2$ both exploit the GSM model of high-quality video frames, but they also have some differences. The S-SpEED features are based on conditional entropies which are weighted by local variances, while VIF uses mutual information between reference and distorted image coefficients. Temporal quality measurements are complementary between the two models: T-SpEED of M$_2$ expresses temporal information loss by conditioning and applying temporal variance weighting, while the TI feature of M$_1$ measures motion changes as a proxy for temporal masking effects.

Interestingly, we found that optimizing weighted averages of the individual predictions from M$_1$ and M$_2$ did not yield significant performance gains. This suggests that the prediction power of the two learners are at near-parity. The prediction averaging step produces a single prediction per frame which is then averaged over all frames of each test video. For the time averaging step, we again employed the hysteresis pooling method \cite{seshadrinathan2011temporal}, as in ST-VMAF.

\section{The VMAF+ Subjective Dataset}
\label{VMAF+}

Data-driven approaches to VQA deeply depend on the training data that is used to train the regressor engines. We believe that a useful training dataset should include a diverse and realistic set of video contents and simulate diverse yet practical distortions of varying degradation levels. Collecting consistent subjective data has the potential to significantly increase the performance of data-driven VQA models on unseen data.

To this end, we conducted a large-scale subjective study targeting multiple viewing devices and video streams afflicted by the most common distortions encountered in large geographic-scale video streaming: compression and scaling artifacts. We first gathered 29 10-second video clips from Netflix TV shows and movies, from a variety of content categories, including, for example, drama, action, cartoon and anime. The source videos were of different resolutions, ranging from 480p up to 1080p, while the frame rates were 24, 25 or 30 frames per second. In our content selection, we also included darker scenes, which are particularly challenging for encoding and video quality algorithms. It should be noted that some of the source videos contain film grain noise. This allows us to gather valuable subjective data on videos that not only suffer from compression and scaling artifacts, but importantly, where there may be degradations of quality in the original source video.

\begin{figure}[htp]
\vspace{-5mm}
\centerline{
\includegraphics[width = 0.875\columnwidth]{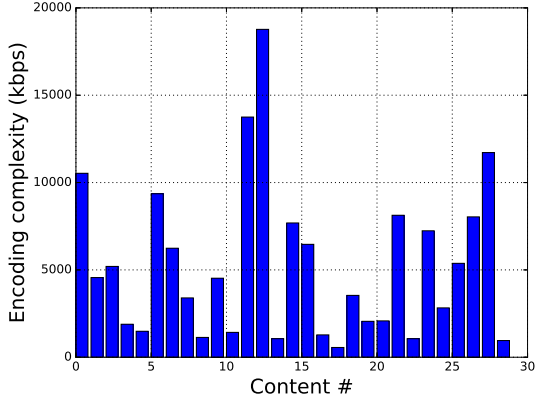}}
\caption{Encoding complexity across contents, expressed as the bitrate (in terms of kbps) of a fixed CRF 23 encode using libx264.}
\label{enc_complexity}
\end{figure}

To describe content variation and encoding complexity, we employed an approach different from the usual SI-TI plots \cite{winkler2012analysis}. We encoded all video contents using a fixed Constant Rate Factor (CRF) setting of 23, then measured the bitrate of each video file. Figure \ref{enc_complexity} shows that the video contents span a large range of encoding complexities, from less than 1Mbps up to around 19 Mbps.

In streaming applications, the source video is usually divided into smaller chunks (e.g. of 2 seconds each) and stored in multiple representations, where each representation is defined by a specific pair of an encoding resolution and bitrate level. To generate the distorted videos, we downsampled each source video to six different encoding resolutions: 320x240, 384x288, 512x384, 720x480, 1280x720 and 1920x1080, then encoded them using the H.264 codec using three different CRF values: 22, 25 and 28, thereby yielding 18 distorted videos per content. For display purposes, all of the videos were upscaled to 1920x1080 display resolution. Both the downscaling and upscaling operations were performed using a lanczos filter. Due to copyright restrictions, the videos cannot be made publicly available.

\begin{figure}[htp]
\vspace{-5mm}
\centering
\includegraphics[width = 0.875\columnwidth]{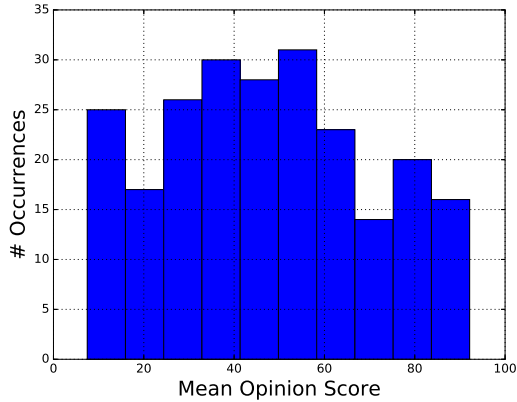}
\caption{Mean opinion score distribution on the VMAF+ database.}
\label{subj_distr}
\end{figure}

To avoid subjective fatigue, we employed a content selection scheme, where each subject only viewed a subset of all video contents. To avoid any memory biases, we ensured that video contents were displayed in a random order such that no video content was consecutively displayed. Overall, we gathered more than 6600 scores from 55 subjects on a laptop viewing device. When training our models, we applied standard subject rejection protocols \cite{BT50013} on the collected human opinion scores. Figure \ref{subj_distr} shows the distribution of the raw subjective data. It can be seen that the scores widely cover the subjective range. The outcome of our subjective test is the VMAF+ video quality database, which we found to be an excellent source of training data for developing learning-based FR-VQA models (see Section \ref{experiments}).

\section{Experimental Analysis}
\label{experiments}

In this section, we discuss a series of experiments on three different and important video quality applications: subjective video quality prediction, Just-Noticeable Difference (JND) prediction and video QoE prediction. For evaluation purposes, we used the Spearman Rank Order Correlation Coefficient (SROCC), the Pearson Linear Correlation Coefficient (PLCC) and the root-mean-squared error (RMSE). The SROCC measures the monotonic relationship between the objective predictions and the ground truth data, while PLCC measures the degree of linearity between the two. The SROCC and PLCC correlation coefficients describe the overall agreement between subjective and objective scores, hence a better objective metric should produce a higher correlation number. Before computing PLCC, a non-linear logistic fitting was applied to the objective scores as outlined in Annex 3.1 of ITU-R BT.500-13 \cite{BT50013}.

We evaluated the proposed approaches against a number of popular FR (and RR) VQA models. We tested the following VQA methods:\footnote{We did not test MOVIE \cite{seshadrinathan2010motion} since it is very time consuming when applied on HD videos.} PSNR, PSNR-hvs \cite{ponomarenko2007between}, SSIM \cite{wang2004image}, MS-SSIM \cite{wang2003multiscale}, ST-RRED \cite{soundararajan2013video}, SpEED-QA \cite{bampis2017speed}, ST-MAD \cite{vu2011spatiotemporal}, VQM-VFD \cite{pinson2014temporal} and VMAF version 0.6.1 \cite{techblog}. For VMAF 0.6.1 we used the suggested parameters, which were also used for E-VMAF. We found this simple parameter selection scheme to work very well for E-VMAF. In our experiments, performing cross-validation for ST-VMAF on the VMAF+ dataset led to overfitting of unseen distortions, hence we empirically fixed $C = 0.5$ and $\gamma = 0.04$. This parameter selection delivered consistent results across a large number of databases, as we will demonstrate next. 

In the experiments, we relied on a wide variety of subjective video databases: LIVE VQA \cite{5404314}, LIVE Mobile \cite{moorthy2012video}\footnote{We excluded frame freezes and used only the mobile subset.}, CSIQ-VQA \cite{vu2014vis3}, NFLX \cite{techblog}, SHVC \cite{shvc_data}\footnote{We excluded videos from Session 3 due to content overlap with the NFLX set.}, VQEG-HD3 \cite{vqeg_hd3}, EPFL-Polimi \cite{de2010h}, USC-JND \cite{wang2017videoset}, LIVE-NFLX \cite{bampis2017study} and LIVE-HTTP \cite{chen2014modeling}. These databases contain a large variety of distortion types, including H.264 and HEVC compression and dynamic rate adaptation, scaling, packet loss, transmission errors and rebuffering events. Importantly, our experimental analysis includes videos with various resolutions, ranging from 352x288 up to 1920x1080, and frame rates (24, 25, 30, 50 and 60 fps). An overview of these databases is given in Table \ref{data_summary}.

\begin{table*}
\caption{Subjective Database Overview. TE: transmission errors, RA: rate adaptation, MJPEG: motion JPEG compression, WC: wavelet-based compression, AWN: additive white noise, QoE: rate adaptation and/or rebuffering. yuv420p8b: planar YUV 420, 8-bit depth, yuv420p10b: planar YUV 420, 10-bit depth.}
\begin{center}
    \label{data_summary}
    \scalebox{0.9825}{
    \begin{tabular}{| c | c | c | c | c | c | c |}
    \hline
    Database & \# Videos & Resolution & Duration & Frame Rate & Format & Distortion Type \\ \hline
    LIVE VQA \cite{5404314} & 150 & 768x432 & 10 sec. & 25, 50 & yuv420p8b & H.264, MPEG-2, TE \\ \hline
    LIVE Mobile \cite{moorthy2012video} & 160 & 1280x720 & 15 sec. & 30 & yuv420p8b & H.264, TE, RA \\ \hline
    CSIQ-VQA \cite{vu2014vis3} & 216 & 832x480 & 10 sec. & \begin{tabular}{@{}c@{}}24, 25, 30 \\ 50, 60\end{tabular} & yuv420p8b & \begin{tabular}{@{}c@{}}H.264, H.265, MJPEG \\ WC, TE, AWN\end{tabular} \\ \hline
    VMAF+ & 290 & 1920x1080 & 10 sec. & 24, 25, 30 & yuv420p8b & H.264 and scaling \\ \hline
    NFLX \cite{techblog} & 300 & 1920x1080 & 6 sec. & 24, 25, 30 & yuv420p8b & H.264 and scaling \\ \hline
    SHVC \cite{shvc_data} & 64 & 1920x1080 & $\cong$ 10 sec. & 25, 50 & \begin{tabular}{@{}c@{}}yuv420p8b \\ yuv420p10b \end{tabular}  & HEVC \\ \hline
    VQEG HD3 \cite{vqeg_hd3} & 135 & 1920x1080 & 10 sec. & 30 & yuv420p8b & H.264, MPEG-2, TE \\ \hline
    EPFL \cite{de2010h} & 144 & \begin{tabular}{@{}c@{}}352x288 \\ 704x576\end{tabular} & 10 sec. & 30 & yuv420p8b & H.264, TE \\ \hline
    USC-JND \cite{wang2017videoset} & 3520 & \begin{tabular}{@{}c@{}}1920x1080, 1280x720 \\ 960x540, 640x360 \end{tabular} & 5 sec. & 24, 30 & yuv420p8b & H.264 \\ \hline
    LIVE-NFLX \cite{bampis2017study} & 112 & 1920x1080 & > 60 sec. & 24, 25, 30 & yuv420p & QoE \\ \hline
    LIVE-HTTP \cite{chen2014modeling} & 15 & 1280x720 & 300 sec. & 30 & yuv420p8b & QoE \\
    \hline
    \end{tabular}
    }
\end{center}
\end{table*}

\subsection{Video Quality Prediction}

\begin{table*}[htp]
\caption{SROCC performance comparison on multiple Video Quality Subjective Databases. VMAF, ST-VMAF and E-VMAF were trained on the VMAF+ dataset. The best overall performance is denoted by boldface.}
\begin{center}
    \label{train_vmaf_plus}
    \scalebox{0.99}{
    \begin{tabular}{| c | c | c | c | c | c | c | c | c | c | c | c |}
    \hline
    Database & LIVE VQA & LIVE Mobile & CSIQ-VQA & NFLX & SHVC & VQEG HD3 & EPFL & \begin{tabular}{@{}c@{}}overall \\ SROCC\end{tabular} & \begin{tabular}{@{}c@{}}overall \\ PLCC\end{tabular} \\ \hline
    PSNR & 0.523 & 0.687 & 0.579 & 0.705 & 0.755 & 0.770 & 0.753 & 0.691 & 0.677 \\ \hline
    PSNR-hvs & 0.662 & 0.757 & 0.599 & 0.819 & 0.828 & 0.798 & 0.904 & 0.785 & 0.788 \\ \hline
    SSIM & 0.694 & 0.757 & 0.698 & 0.788 & 0.754 & 0.907 & 0.712 & 0.771 & 0.752 \\ \hline
    MS-SSIM & 0.732 & 0.748 & 0.749 & 0.741 & 0.715 & 0.898 & 0.931 & 0.808 & 0.791 \\ \hline
    ST-RRED & 0.805 & 0.892 & 0.805 & 0.764 & 0.889 & 0.912 & 0.944 & 0.872 & 0.777 \\ \hline
    SpEED-QA & 0.776 & 0.897 & 0.741 & 0.781 & 0.879 & 0.909 & 0.936 & 0.861 & 0.759 \\ \hline
    ST-MAD & 0.825 & 0.663 & 0.735 & 0.768 & 0.611 & 0.847 & 0.901 & 0.782 & 0.769 \\ \hline
    VQM-VFD & 0.804 & 0.816 & 0.839 & 0.931 & 0.863 & 0.939 & 0.850 & 0.873 & 0.870 \\ \hline
    VMAF 0.6.1 & 0.756 & 0.906 & 0.614 & 0.928 & 0.887 & 0.850 & 0.836 & 0.847 & 0.853 \\ \hline
    ST-VMAF & 0.809 & 0.905 & 0.784 & 0.927 & 0.888 & 0.932 & 0.945 & \textbf{0.897} & \textbf{0.898} \\ \hline
    E-VMAF &  0.792 & 0.929 & 0.761 & 0.930 & 0.892 & 0.906 & 0.942 & 0.894 & 0.895 \\
    \hline
    \end{tabular}
    }
\end{center}    
\end{table*}

We begin our experimental analysis with the problem of video quality prediction. To accurately evaluate performance, we focused on cross-database results, i.e., we relied on the VMAF+ subjective dataset for training and tested on the rest of the video databases. For each VQA model, we report the SROCC values per testing dataset, as well as an aggregate SROCC and PLCC value. To compute the aggregate correlation score, we applied Fisher's z-transformation \cite{corey1998averaging}, i.e., 

\begin{equation}\label{fisher}
z=\frac{1}{2}\mathrm{ln}\frac{1+r}{1-r}, \mathrm{where} \ r \ \mathrm{is \ SROCC \ or \ PLCC,}
\end{equation}
to the correlation values and then averaged them over all tested databases. The average value was then transformed back using the inverse of \eqref{fisher}. Table \ref{train_vmaf_plus} shows the results of this experimental analysis.

Among image-based models, such as PSNR and SSIM, PSNR delivered the worst performance. This is expected, since it is a signal fidelity metric that does not exploit perceptual information. SSIM and PSNR-hvs performed considerably better and MS-SSIM achieved further performance gains, likely due to the multiscale properties captured therein. Nevertheless, none of these spatial metrics exceeded an aggregate SROCC of 0.81, which demonstrates the importance of capturing temporal information.

Regarding video-based models, ST-MAD did not perform well and was very time-consuming (see Section \ref{computational_analysis}). VMAF 0.6.1 delivered excellent performance on the NFLX dataset, which is expected, given that it mostly captures compression and scaling artifacts. However, it demonstrated poor generalization capabilities on unseen distortions, such as the CSIQ-VQA database. ST-RRED and SpEED-QA performed well on most databases in terms of SROCC, but neither algorithm performed well on the NFLX dataset, which may be due to the presence of film grain in some of the source content. Notably, the aggregate PLCC of ST-RRED and SpEED-QA was relatively low. Unlike VQA models trained on subjective data, such as VMAF or VQM-VFD, the ST-RRED and SpEED-QA predictions were highly non-linear with ground truth. VQM-VFD delivered similar SROCC performance, but, unlike ST-RRED and SpEED-QA, it uses a number of basic features that are fed to a neural network trained on a very large number of subjective datasets.

From the above analysis, it can be seen that VMAF does not fully exploit temporal information and does not generalize well on unseen distortions. At the same time, untrained VQA models such as ST-RRED do not exhibit a linear relationship with subjective ground truth, do not capture the effects of film grain and do not combine multiscale information. The methods we have developed here aim to bridge this gap and combine the best of both worlds. Table \ref{train_vmaf_plus} shows that ST-VMAF achieved standout aggregate performance across all databases, while E-VMAF functioned nearly as well. Both models achieve this excellent level of video quality prediction power using a \textit{single training dataset and a single parameter setting}. It should be noted that both ST-VMAF and E-VMAF considerably improve on VMAF, although they were trained on the VMAF+ dataset, which focuses only on compression and scaling artifacts. For example, on the CSIQ-VQA database, which contains multiple distortion types other than compression, ST-VMAF and E-VMAF both perform quite well. This strongly suggests that these new models possess excellent generalization capabilities beyond their demonstrated state-of-the-art VQA performance. In the Appendix, we further analyze the cross-database performances and the computational complexities of the proposed VQA models.

\subsection{Monotonicity Analysis}

As already discussed, E-VMAF does not require an increased feature dimensionality as ST-VMAF does, and hence, is less likely to overfit. To demonstrate the effects of overfitting, we studied the prediction consistency of ST-VMAF and E-VMAF when the encoding resolution and the compression ratio were varied. Given any two compression levels $c_1\le c_2$, a consistent objective model should satisfy $o_{c_1}\le o_{c_2}$, where $o_{c_1}$ and $o_{c_2}$ are the respective calculated quality predictions. Likewise, for any two encoding resolutions $r_1\le r_2$, then $o_{r_1}\le o_{r_2}$ should also be satisfied. A VQA model should satisfy these two ``monotonicity" properties. If a trained model is not monotonic with compression and resolution, it may be an indication of overfitting. Another benefit of a monotonic VQA model is that it can be used for codec comparisons and perceptual video encoding.

\begin{figure*}[htp]
\centerline{
\includegraphics[width = 0.65\columnwidth]{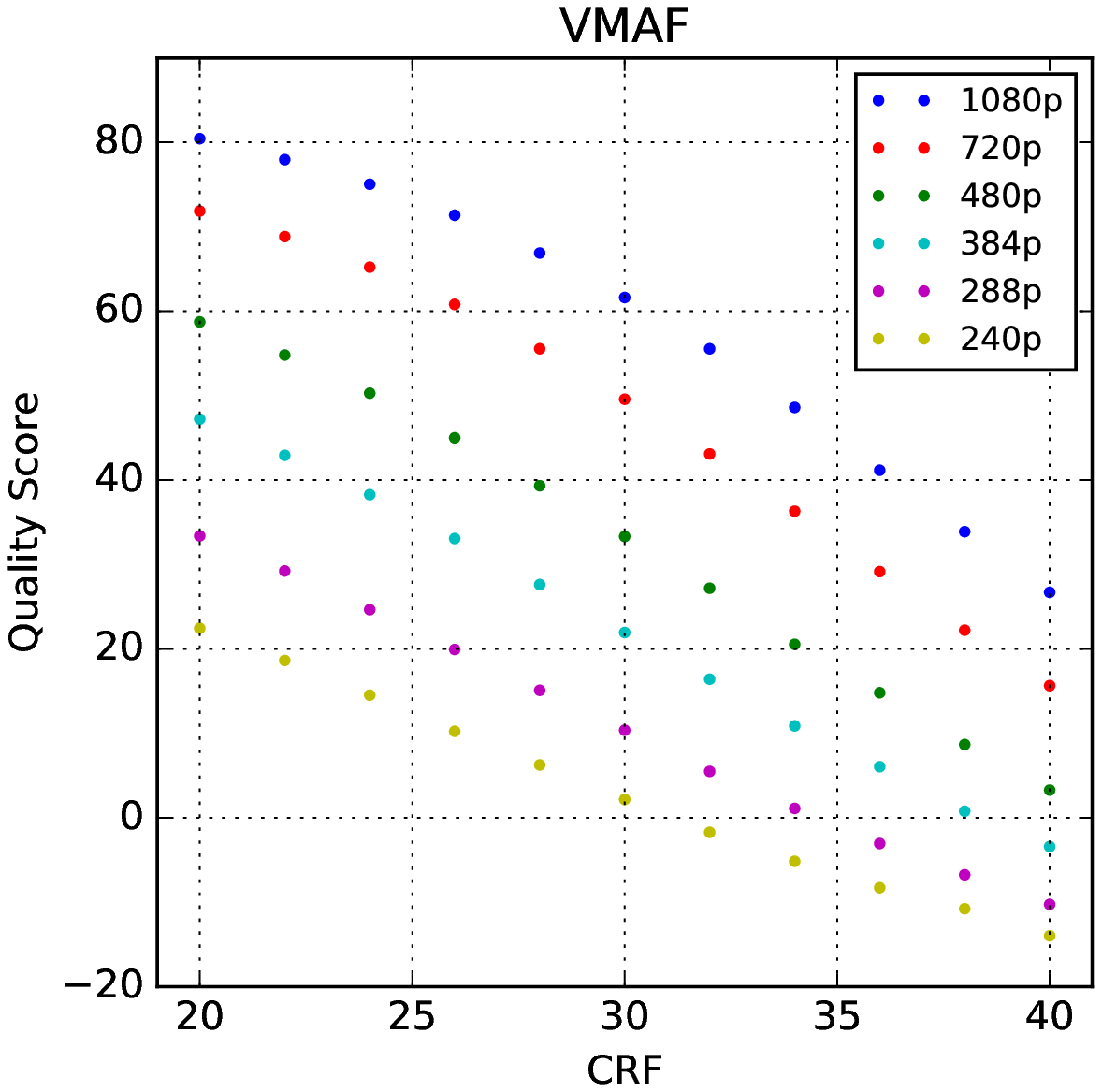}
\includegraphics[width = 0.65\columnwidth]{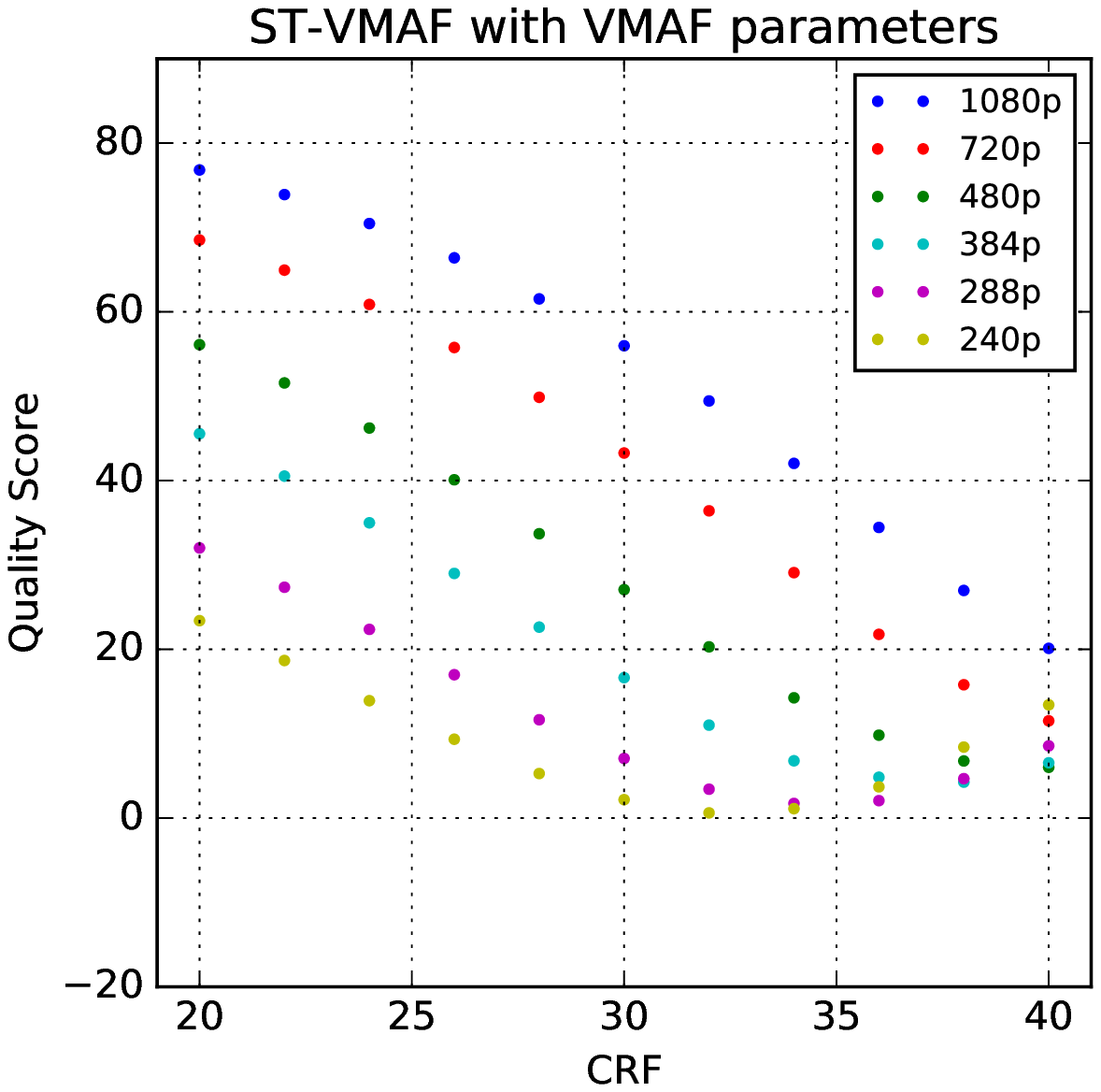}
\includegraphics[width = 0.65\columnwidth]{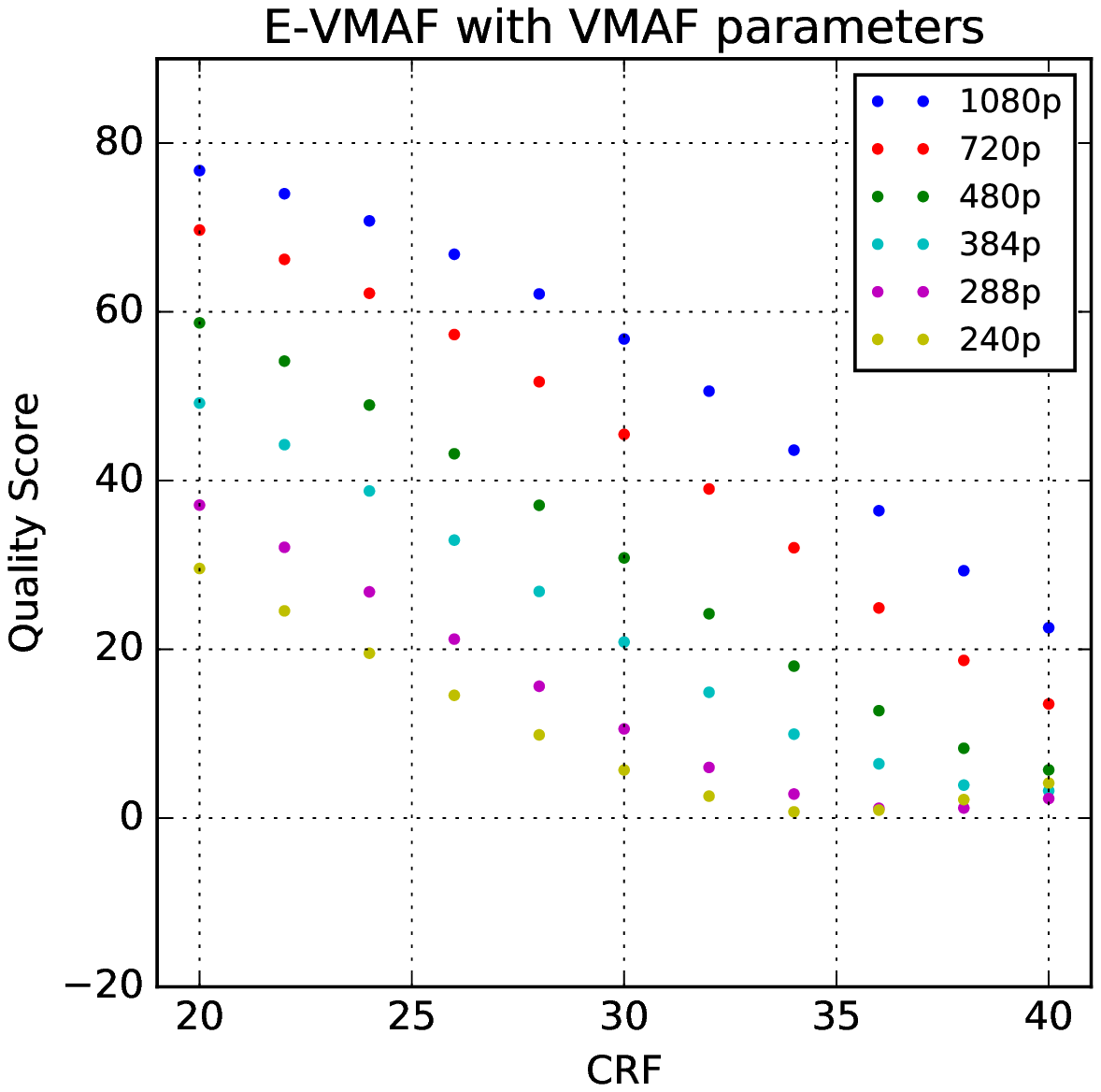}
}
\caption{Monotonicity plots for VMAF variants. Left: VMAF 0.6.1; Center: ST-VMAF; Right: E-VMAF.}
\label{stevmaf_monotonicity}
\end{figure*}

To study the monotonicity of ST-VMAF and E-VMAF, we conducted the following experiment. First, we selected a 1080p video content from the NFLX set. This particular video segment is an action scene of 10 sec. (240 frames) duration and is challenging in terms of preserving monotonicity. We encoded it at six different resolutions: 1080p, 720p, 480p, 384p, 288p and 240p, and at 11 constant rate factor (CRF) values, ranging from 20 to 40 in steps of 2. When the original video was encoded at a resolution less than 1080p, it was upscaled to 1080p, before measuring video quality. The underlying assumption was that scaling artifacts at lower encoding resolutions should be monotonically captured by the video quality predictor. As shown in Fig. \ref{stevmaf_monotonicity}, both ST-VMAF and E-VMAF are nicely monotonic over most of the compression range when using the default VMAF 0.6.1 parameters ($C=4$ and $\gamma=0.04$).

However, for large CRF values (more severe compression artifacts), ST-VMAF loses its monotonic behavior across resolutions, which may be indicative of overfitting due to its larger feature dimensionality as compared to VMAF 0.6.1. By contrast, using the same parameters, E-VMAF better preserves monotonicity, although there is still some room for improvement. VMAF 0.6.1 has been tuned to retain its monotonic behavior and hence, in practice, both ST-VMAF and E-VMAF need to be further manually tuned for monotonicity-optimal results.

\subsection{JND Prediction}

Another interesting application of VQA models is JND detection, i.e., identifying JND points, and comparing them with the detection capabilities of humans. The USC-JND dataset \cite{wang2017videoset} was designed specifically for this purpose. It contains 220 video contents encoded at four resolutions (1080p, 720p, 540p and 360p) and multiple quantization parameter (QP) values. For each content and resolution, the QP values corresponding to the first 3 JND points are determined via the following procedure. First, a video encoded with QP 0 is used as an anchor point in a binary search procedure that determines the QP value for the first JND point. This QP value is subsequently used to determine the QP value for the second JND point, and so on. Along with the ``anchor" point (QP 0), this process yields $220\times4\times4=3520$ videos labeled by subjective scores 0, 1, 2 and 3 for each initial (anchor) point and 3 associated JND points.

To evaluate the JND detection capabilities of the proposed models, we selected several leading VQA models and reported their JND prediction performance in Table \ref{usc-jnd}. For this detection task, we did not employ the hysteresis temporal pooling, since detection is a different task. Both ST-VMAF and E-VMAF outperformed other powerful VQA models, including ST-RRED and VMAF.

\begin{table}[htp]
\caption{USC-JND performance comparison. VMAF, ST-VMAF and E-VMAF were trained on the VMAF+ dataset. The best performing algorithms are denoted by boldface.}
\begin{center}
\label{usc-jnd}
\begin{tabular}{| c | c | c | c | c |}
\hline
Database & SROCC & PLCC \\ \hline
PSNR & 0.616 & 0.589 \\ \hline
SSIM & 0.718 & 0.602 \\ \hline
MS-SSIM & 0.815 & 0.739 \\ \hline
ST-RRED & 0.844 & 0.735 \\ \hline
SpEED-QA & 0.843 & 0.727 \\ \hline
VMAF 0.6.1 & 0.853 & 0.854 \\ \hline
\textbf{ST-VMAF} & \textbf{0.877} & \textbf{0.856} \\ \hline
\textbf{E-VMAF} & \textbf{0.875} & \textbf{0.869} \\
\hline
\end{tabular}
\end{center}    
\end{table}

\subsection{QoE Prediction}

An important emerging application of perceptual video quality models is streaming video QoE prediction. In streaming applications, the reference video is usually available, hence reference models are more relevant. The predominant video impairments that occur during video streaming are compression, spatial scaling artifacts, and rebuffering events. We studied the behavior of the ST-VMAF and E-VMAF models on the recently released LIVE-NFLX Video QoE Database \cite{bampis2017study}, which simulates realistic buffer and network constraints, and contains rebuffering events, rate adaptations and constant bitrate encodes. Table \ref{LIVE_NFLX} shows that none of the considered FR-VQA models performed particularly well, since they do not model the effects of rebuffering. This suggests that more sophisticated QoE predictors (than just VQA algorithms) are required for the more general problem of QoE assessment. However, both of the new models achieved better performance than all the other models, especially in terms of PLCC\footnote{VQM-VFD cannot be applied to videos of duration more than 15 sec. and hence is excluded.}. 

\begin{table}[htp]
\caption{Quantitative performance comparison on the  LIVE-NFLX Video QoE Database \cite{bampis2017study}, including both compression and rebuffering events. The best performing algorithm is denoted by boldface.}
\begin{center}
\label{LIVE_NFLX}
\begin{tabular}{ |c|c|c|c| }
\hline
VQA & SROCC & PLCC \\ \hline
PSNR & 0.515 & 0.507 \\ \hline
PSNR-hvs & 0.535 & 0.545 \\ \hline
SSIM & 0.701 & 0.726 \\ \hline
MS-SSIM & 0.683 & 0.710 \\ \hline
ST-RRED & 0.702 & 0.715 \\ \hline
SpEED-QA & 0.712 & 0.727 \\ \hline
VMAF 0.6.1 & 0.607 & 0.667 \\ \hline
\textbf{ST-VMAF} & \textbf{0.735} & \textbf{0.780} \\ \hline
E-VMAF & 0.721 & 0.772 \\
\hline
\end{tabular}
\end{center}    
\end{table}

We also examined the potential of incorporating the ST-VMAF and E-VMAF VQA models into an existing continuous-time QoE predictor. We tested the revised QoE predictor using the LIVE-HTTP Video QoE Database \cite{chen2014modeling} which studies the effects of HTTP-based rate adaptation on 5 min. long HD video sequences. Table \ref{NARX_Chao} reports the outcomes of the experiments when using the NARX QoE predictor \cite{7931662}, which has demonstrated promising results on the few available video QoE databases. First, we split the database into content independent train-test splits, then determined the best NARX configuration on the training set. Next, we tested the selected parameter setting on the test videos using a number of leading VQA models as integral components of the NARX QoE predictor. For evaluation purposes, we reported the SROCC and root mean squared error (RMSE) values between the continuous QoE predictions and the continuous ground truth data. It can be seen that ST-VMAF outperformed all of the other VQA models when used in this way, suggesting that it is an excellent choice for inclusion in future perceptually-driven online QoE prediction systems. Ultimately, we envision deploying high-performance QoE predictors to design practical perception-driven rate adaptation and network allocation protocols.

\begin{table}[htp]
\caption{Quantitative performance comparison on the LIVE-HTTP \cite{chen2014modeling} Video QoE Database when using the continuous-time NARX \cite{7931662} QoE predictor. The best performing algorithm is denoted by boldface.}
\begin{center}
\label{NARX_Chao}
\begin{tabular}{ |c|c|c|c| }
\hline
VQA & SROCC & RMSE \\ \hline
PSNR & 0.731 & 6.708 \\ \hline
SSIM & 0.901 & 3.844 \\ \hline
MS-SSIM & 0.881 & 4.248 \\ \hline
ST-RRED & 0.885 & 4.226 \\ \hline
VMAF 0.6.1 & 0.883 & 4.321 \\ \hline
\textbf{ST-VMAF} & \textbf{0.924} & \textbf{3.515} \\ \hline
E-VMAF & 0.922 & 3.666 \\
\hline
\end{tabular}
\end{center}    
\end{table}

\subsection{Observations and Takeaways}
In our experiments, we demonstrated that both ST-VMAF and E-VMAF performed very well for video quality and JND prediction and have the potential to be integrated with QoE predictors. Between the two, their performances are quite similar: E-VMAF was slightly better in terms of PLCC for JND prediction (see Table \ref{usc-jnd}) and ST-VMAF was a bit better in terms of SROCC in the LIVE-NFLX experiment (see Table \ref{LIVE_NFLX}). The main benefit of using E-VMAF is that it better preserves monotonicity at very high compressions (see Fig. \ref{stevmaf_monotonicity}). Also, it is easier to tune, since using the same SVR parameters as VMAF yielded excellent results. By contrast, to train ST-VMAF, its larger number of features (compared to the VMAF baseline) had to be regularized using more careful SVR tuning. Nevertheless, in applications where a compact feature and model representation is required, ST-VMAF might be a preferred solution.

\section{Future Work}
\label{Future}

We developed two high-performing, data-driven full reference video quality assessment models. In the future, we plan to further improve those models by combining NR source VQA measurements with the FR system towards accounting for possible degradations of the original source/reference video. To do so, we also plan to develop better data-driven NR video quality models that can be used in lieu of existing NR VQA \cite{saad2014blind} approaches. Towards achieving this goal, it will be very interesting to exploit the ensemble fusion idea proposed here on the NR VQA problem.

\section{Acknowledgements}
The authors would like to acknowledge Anush K. Moorthy, Ioannis Katsavounidis and Anne Aaron for their support.

\section{Appendix}
\subsection{Cross-database performance for ST-VMAF and E-VMAF}
\label{other_train_db_analysis}

The proposed models rely on three components: the VMAF+ training subjective data, the spatiotemporal feature integration and the temporal pooling step. In this section, we investigate the effects on the predictive performance of ST-VMAF and E-VMAF when each of these components varies.

First, we investigated the effects on the predictive performance of ST-VMAF when trained on other databases in Table \ref{cross_stvmaf}. Importantly, the VMAF+ dataset proved to be highly consistent, and served as an excellent training dataset for ST-VMAF. It is also encouraging that the aggregate SROCC values (for a fixed training dataset) achieved by ST-VMAF were very close to, or significantly exceeded 0.8 (see second to last column in Table \ref{cross_stvmaf}). Similar observations apply to the E-VMAF predictions. 

\begin{table*} [htp]
\caption{Cross-database SROCC for ST-VMAF. Each element in this matrix shows the SROCC performance when training on the dataset in the row and testing on the dataset in the column. The last two columns show the aggregate SROCC and PLCC performance per training dataset. Using the VMAF+ dataset for training yielded the best overall performance and is denoted by boldface.}
\begin{center}
\label{cross_stvmaf}
    \scalebox{0.9}{
    \begin{tabular}{| c | c | c | c | c | c | c | c | c | c | c | c |}
    \hline
    Database & LIVE VQA & LIVE Mobile & CSIQ-VQA & VMAF+ & NFLX & SHVC & VQEG-HD3 & EPFL & \begin{tabular}{@{}c@{}}overall \\ SROCC\end{tabular} & \begin{tabular}{@{}c@{}}overall \\ PLCC\end{tabular} \\ \hline
    LIVE VQA & - & 0.869 & 0.742 & 0.775 & 0.859 & 0.873 & 0.628 & 0.842 & 0.811 & 0.809 \\ \hline
    LIVE Mobile & 0.584 & - & 0.736 & 0.791 & 0.900 & 0.891 & 0.653 & 0.836 & 0.794 & 0.786 \\ \hline
    CSIQ-VQA & 0.599 & 0.856 & - & 0.757 & 0.855 & 0.879 & 0.669 & 0.830 & 0.795 & 0.794 \\ \hline
    VMAF+ & 0.809 & 0.905 & 0.784 & - & 0.927 & 0.888 & 0.932 & 0.945 & \textbf{0.897} & \textbf{0.898} \\ \hline
    NFLX & 0.733 & 0.925 & 0.754 & 0.888 & - & 0.874 & 0.922 & 0.947 & 0.882 & 0.884 \\ \hline
    SHVC & 0.700 & 0.893 & 0.759 & 0.808 & 0.866 & - & 0.887 & 0.930 & 0.850 & 0.846 \\ \hline
    VQEG-HD3 & 0.706 & 0.890 & 0.732 & 0.813 & 0.879 & 0.822 & - & 0.933 & 0.842 & 0.839 \\ \hline
    EPFL & 0.715 & 0.931 & 0.717 & 0.866 & 0.918 & 0.878 & 0.879 & - & 0.862 & 0.859 \\
    \hline
    \end{tabular}
    }
\end{center}    
\end{table*}

Having established that training on VMAF+ is the best option, we studied how the performance of ST-VMAF and E-VMAF compares to that of the individual models M$_\mathrm{1}$ and M$_\mathrm{2}$, and the performance gains of hysteresis pooling. To this end, we report the results (when training on VMAF+) in Table \ref{cross_db}. It can be observed that M$_\mathrm{1}$ and M$_\mathrm{2}$ (Section \ref{EVMAF}) deliver similar performances, but afford significant performance gains when combined using E-VMAF. Similarly, ST-VMAF combines some features that may also belong to either M$_\mathrm{1}$ or M$_\mathrm{2}$, but their combination performs significantly better. Hysteresis pooling further improves the predictive performance of both ST-VMAF and E-VMAF.

\begin{table}[htp]
\caption{Cross-database Aggregate Performance(training on VMAF+ dataset). The best performance is denoted by boldface.}
\begin{center}
\label{cross_db}
    \scalebox{1}{
    \begin{tabular}{| c | c | c | c | c | c |}
    \hline
    Database & pooling & SROCC & PLCC \\ \hline
    M$_\mathrm{1}$ & mean & 0.847 & 0.853 \\ \hline
    M$_\mathrm{2}$ & mean & 0.845 & 0.847 \\ \hline
    ST-VMAF & mean & 0.885 & 0.887 \\ \hline
    E-VMAF & mean & 0.873 & 0.875 \\ \hline
    \textbf{ST-VMAF} & hysteresis & \textbf{0.897} & \textbf{0.898} \\ \hline
    E-VMAF & hysteresis & 0.894 & 0.895 \\
    \hline
    \end{tabular}
    }
\end{center}    
\end{table}

\subsection{Computational Analysis for ST-VMAF and E-VMAF}
\label{computational_analysis}

\begin{figure}[htp]
\centering
\vspace{-4mm}
\includegraphics[width = 0.85\columnwidth]{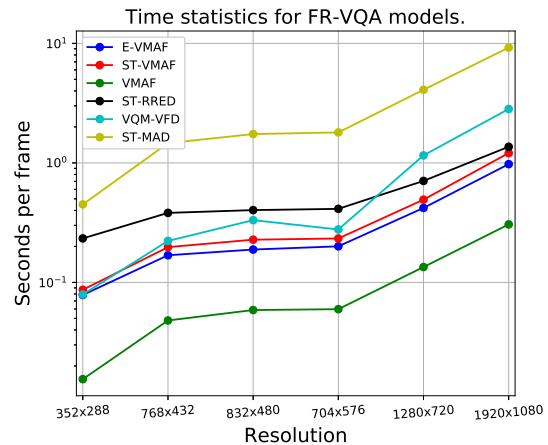}
\caption{Per frame compute time required for each FR-VQA model (log vertical scale).}
\label{time_stats}
\end{figure}

To deploy VQA models for video quality prediction at global scale, ensuring low time complexity is a critical requirement. Therefore, we studied the per frame compute time consumed by several leading FR-VQA models\footnote{Frame-based models are usually much faster and hence are excluded.} in Figure \ref{time_stats}. For our analysis, we selected videos from 6 different resolutions ranging from CIF (352x288) up to Full HD (1920x1080). These videos have 334 frames on average and we averaged our time calculations over 5 trials. All of the compute time analysis was carried out on a 16.04 Ubuntu LTS Intel i7-4790@3.60GHz system.

ST-MAD required the most compute time, followed by ST-RRED and VQM-VFD. When implementing ST-MAD and VQM-VFD, we encountered out-of-memory issues on long Full HD videos. This could be due to the fact that the ST-MAD implementation stores and loads the entire video into memory. Another limitation of these approaches is that they only process entire videos with no capability to produce continuous video quality scores. ST-RRED processes the video frame by frame to produce continuous scores, but requires calculating a complete multi-scale, multi-orientation steerable decomposition. 

By contrast, ST-VMAF and E-VMAF are memory efficient, produce continuous quality scores and consume less compute time, since they extract the very efficient S-SpEED and T-SpEED features. Our ST-VMAF and E-VMAF implementation uses un-optimized Matlab code to extract SpEED-QA features, while VMAF uses AVX optimization and is implemented in C. Since ST-VMAF and E-VMAF are natural extensions within the VMAF ecosystem, it is possible to adopt similar optimization approaches.


\bibliographystyle{IEEEtran}
\bibliography{bibfile_stevmaf}{}

\end{document}